\documentclass[12pt,a4paper]{article}

\usepackage{graphicx}
\usepackage{dirtytalk}
\usepackage{multirow}
\usepackage{hyperref}
\usepackage{natbib}
\usepackage{amsmath}
\usepackage[nopatch]{microtype}
\usepackage{booktabs}
\usepackage{setspace}
\begin{document} 

\title{Gender Dynamics in Russian Online Political Discourse}
\author{}
\date{}

\maketitle 

\author{Elizaveta Savchenko\(^1\) and Michael Raphael Freedman\(^{2,*}\) \\  \\\(^1\) Department of Mathematics, Ariel University \\ 
\(^2\) Department of Political Science, University of Haifa\\ 
\(^*\) Corresponding author: mrfreed@poli.haifa.ac.il}

\begin{abstract}
\noindent The digital landscape provides a dynamic platform for political discourse, crucial for understanding shifts in public opinion and engagement, especially under authoritarian governments. This study examines YouTube user behavior during the Russian-Ukrainian war, analyzing 2,168 videos with over 36,000 comments from January 2022 to February 2024. We observe distinct patterns of participation and gender dynamics that correlate with major political and military events. Notably, females were more active in anti-government channels, especially during peak conflict periods. Contrary to assumptions about online
engagement in authoritarian contexts, our findings suggest a complex interplay where women emerge as pivotal digital communicators. This highlights online platforms’ role in facilitating political expression under authoritarian regimes, demonstrating its potential as a barometer for public sentiment. \\

\textbf{Keywords}: digital activism, social natural language processing, online political discourse, Youtube.

\end{abstract}

\newpage
\section{Introduction}
\doublespacing
The advent of the digital revolution has democratized information access and enabled unparalleled levels of connectivity, empowering individuals to engage in political discourse independently of conventional gatekeepers \citep{17,1}. Today, the internet presents significant opportunities for engaging large populations in political involvement, even in non-democratic states \citep{1,2}.  Specifically, Internet usage offers anonymity and the ability of non-physical presence in the political information space \citep{4,5}. This dynamic complicates the efforts of authoritarian regimes to suppress dissent compared to traditional offline methods of detaining political activists \citep{7}. Hence, the internet fosters freedom of expression, opinion, and facilitates communication, thereby influencing the development of weak social ties that can lead to collective political action in the future \citep{8}.

According to \cite{9}, in democratic states, government officials and political figures widely utilize the Internet to promptly update citizens on the political landscape and foster communication between elected representatives and the public. However, in authoritarian regimes, the Internet serves different purposes. Authoritarian governments utilize the Internet and social media platforms to disseminate propaganda, spread disinformation, discourage citizens from engaging in political activities, employ intimidation tactics, and conduct surveillance \cite{11,12}. In non-democratic countries, Internet resources make it possible to create opposition movements, thereby introducing a new element that alters the dynamics and offers optimism for attaining freedom \citep{10}. Low barriers to entry on Internet networks, compared to traditional media, make it possible to enter the public space without intermediaries and quickly adapt to the pressure of state censorship, which, in turn, can potentially make political regimes more vulnerable \citep{13,14}. 

For instance, using the example of the 2009 Iranian elections, \cite{14a} and \cite{14b} found that the main utility of Twitter was not only organization/mobilization or action/reaction, but in the category of awareness/advocacy, particularly with respect to the international audience. The other side of this phenomenon is the exposure of diverse perspectives through accessible information resources can have its drawbacks, such as a rise in populism, the proliferation of fake news for increased viewership, disinformation stemming from unethical journalism, or the dissemination of extremist ideologies \citep{15}. 

The capabilities offered by the Internet have emerged as a significant factor in shaping civic awareness in Russia during the 21st century \citep{16}. For instance, \cite{16} highlighted the correlation between the extensive use of the Russian social network Vkontakte and involvement in the 2011 protests in Russia. Specifically, it was observed that the reduction in costs and resources required for preparatory communication facilitated the engagement of a larger number of individuals in offline protest activities. In a more general sense, platforms like Vkontakte, Facebook, X (formally, Twitter), Telegram, Instagram, and YouTube have emerged as virtual town halls where diverse voices converge to discuss pressing socio-political issues, shaping public opinion and influencing policy agendas \citep{18}. The widespread use of the Internet in Russia has greatly undermined television as a source of socially significant political news and analytical programs \citep{19}. To be precise, as of 2021, television is the main source of information for 17\% of Russians, mainly in the 50+ age group. At the same time, 90\% of the population has access to the Internet \footnote{\url{https://datahub.itu.int/data/?e=RUS&c=701&i=11624}, \url{https://datareportal.com/reports/digital-2023-russian-federation}}. Significantly, approximately 80\% of Russian citizens aged 18 to 59 years old say that they use the Internet as a source of information, but do not exclude other methods of distribution \citep{20}\footnote{\url{https://wciom.ru/analytical-reviews/analiticheskii-obzor/mediapotreblenie-rossijan-monitoring}}. 

Nonetheless, a significant portion of Russian citizens show little interest in engaging with political content online \citep{21a,21b}. Decades of Soviet and Putin-era propaganda promoting \textbf{apoliticism} and political apathy have effectively discouraged a large segment of the Russian populace from active political participation \citep{21,22,23}. However, recent events, including the Russian-Ukrainian war, have catalyzed increased citizen engagement in the country's political landscape \citep{20}. As of 2024, political discourse on the internet is de facto prohibited \citep{26}. In such repressive circumstances, political expression online is viewed as an act of civil opposition to the government \citep{26a}.

In the context of the \textit{de facto} abolition of freedom of speech, freedom of assembly, and the criminalization of political activity, it is difficult to draw correct conclusions about the state of Russian society through the articulation of interests \citep{27,28}. Taken jointly with the fact that, unlike direct democracies, Russia is a hybrid personalist autocracy or according to other estimates, a full-fledged personalist dictatorship, so public demands can be measured only by indirect data and signs \citep{30,31}. For instance, using implicit participation in political activities online - like comments on videos associated with one or another part of the political position in Russia. 

The demographic distribution of politically active individuals holds significance \citep{32,33,34,35,36}. Political preferences and demands, such as the age, gender, educational attainment, and marital status of protesters; local protest customs; the extent of state repression; and numerous other factors, influence the selection of protest strategies and the outlook for protest activity overall \citep{37,38,39}. In addition, global trends in the protest strategy preferences of women and men, as well as their attitudes toward the effectiveness and appropriateness of various forms of social resistance and expression of will, are noteworthy \citep{33,36,41a,41}. Understanding the demographic peculiarities of each society is crucial for drawing conclusions regarding the likelihood, level of violence, and content of protest demands \citep{42,43,44,45,46}. Furthermore, diverse socio-economic and cultural elements impact an individual's political stance, which influences methods of expressing a civil political stance, and consequently, their affiliation with a political party or non-systemic political movement \citep{47}. These and other factors, to varying degrees, are the cause of the polarization of society, which also manifests itself in Internet activities, among other things \citep{48}. 

Political polarization in online spaces, often referred to as \say{echo chambers}, can exacerbate existing divisions by creating information bubbles where users are shielded from opposing viewpoints and consistently exposed only to information that aligns with their worldview \citep{49,50}. Concurrently, social media platforms can paradoxically lead to reduced face-to-face communication on socio-political issues. Certain behaviors on these platforms are linked to decreased perceived agreement of opinions among social connections. This, in line with the spiral of silence theory, contributes to a decreased willingness among social media users to engage in political conversations offline in certain contexts \citep{51}.

Notwithstanding the challenges inherent in undertaking a traditional opinion survey, it remains feasible to discern predominant political tendencies by analyzing the online activities of Russian citizens. Thus, one can focus on social signals and online activities as content consumption and participation in such context. Consequently, there's a shift toward examining social cues and online activities such as content consumption and participation through comments within this framework. Since the mid-2010s, the Russian YouTube community has experienced significant politicization, particularly among individuals under 40 years old \citep{53}. As of 2023, YouTube stands as the leading video platform in Russia \footnote{\url{https://mediascope.net/data/}}. Despite being subject to Russian regulations, more often than not, YouTube continues to uphold freedom of speech. Inevitably, there is a trend in Russia towards co-opting the online information space by pro-government entities to propagate their narratives, especially given the limited influence of the state in this domain. In addition, YouTube channels were present before the beginning of the Russian-Ukrainian war, unlike other popular platforms such as Telegram, allowing to explore the pre-war sentiment.  

In this study, we utilize state-of-the-art artificial intelligence (AI) models that are able to detect the gender of a YouTube account according to their username and comment’s content, in Russian. Using this capability, we collected data from 30 popular Russian-speaking YouTube channels from the beginning of the Russian-Ukrainian war in January 2022 and up to February 2024. Half of the channels are associated with the government and its regime while the other half is associated with the opposition. Using this data, we explore trends in active participation in the political informal life and its unique dynamics during the war with respect to gender-based differences as a demographic indicator of the participants. We find that woman are much more active on anti-government channels, and that they are more likely to comment during times of heightened conflict.

This paper makes several significant contributions to the understanding of online political discourse in authoritarian contexts, particularly focusing on gender dynamics during the Russian-Ukrainian war. By employing advanced AI models to analyze YouTube comments on Russian political channels from January 2022 to February 2024, the study provides novel insights into gender-based patterns of online political participation, revealing increased female engagement during periods of heightened conflict. The research identifies significant shifts in discourse corresponding to major war events, demonstrating digital platforms' responsiveness to real-world developments. Through comparative analysis of pro-government and anti-government channels, the study also offers insights into how political alignment influences engagement patterns. The methodological innovation of using AI for gender prediction based on usernames and comment content represents a novel approach in studying online political discourse. By tracking the evolution of political narratives through content analysis, the study contributes to our understanding of how digital activism functions in restrictive political environments. This interdisciplinary approach, bridging political science, digital communication studies, and gender studies, provides a multifaceted perspective on online political engagement and opens new avenues for research into online political behavior, gender roles in digital activism, and the impact of social media on public opinion in authoritarian contexts.

% The rest of the paper is organized as follows. The next section outlines the theoretical framework used to explore the gender composition of Russian-speaking participants in online communications on political and socially significant topics. In addition, we formally define the proposed AI model’s development and usage as well as the data obtained by it. Next, we report the trends and dynamics revealed by the obtained data. Finally, we discuss the results in a socio-demographic context and suggest possible future studies.

\section{Methods and Materials}
In this section, we outline our data collection procedure and statistical analysis. 

%\subsection{Theoretical framework}

%Is commenting on YouTube videos a form of political participation? 
%Are people worried about censorship?
%Are the comments made anonymously?
%Why would the war make women more likely to get politically involved?

%(a lot of the research focuses on the regime’s perspective: potential benefits in boosting women’s representation or rights). Less on the motivations for women to participate?

%Digital Activism in Russia: The Evolution and Forms of Online Participation in an Authoritarian State Markku Lonkila, Larisa Shpakovskaya, and Philip Torchinsky

\subsection{Data collection}
In order to collect data for this study, we used the Python programming language (version 3.9.2) \citep{python}. In particular, we used the Selenium library \citep{selenium} to explore all the videos in each of a pre-defined list of channels (see Table \ref{table:appendix}). Each channel is associated with either a pro- or anti- government position. Therefore, we assume each video in a channel is also associated with the same position. 

For each video, we collected the video’s publication date which is relative to the time when the data was collected, and all the comments in terms of the user’s name, the user’s comment, and the comment's date. Importantly, comments answered to other comments are excluded from this study as it is assumed that these refer to the user's generated content rather than the video's content. 

\subsection{Content Analysis and Predicted Gender}
In order to associate a gender with a user, we used a two-phase approach. First, we used the model proposed by \citep{gender_model} which was trained on over 100 million pairs of names and gender associations, as collected by Yahoo. The authors of the model report over 97\% accuracy on the validation cohort. For our case, a user's gender is allocated only if the model's prediction confidence is higher than 90\%. If this model’s confidence was not high enough, due to the usage of non-personal name-related usernames or any other reason, we used the model developed by \cite{gender_model} to differentiate the user’s gender based on their comment text. Research shows that a Multi-layered perception model, a relatively simple version of the neural network model can differentiate between genders trained on Twitter text, which has a 280-character limit (at the time of the analysis), with 82\% accuracy in the validation cohort \citep{dl_in_nlp}. Since the comments text we analyzed are also relatively short, we have chosen this model. If this model produces a prediction with at least 90\% confidence, we allocate the predicted gender to the user. 

\subsection{Statistical analysis}
Based on the collected data, we computed the gender distribution over time and for pro- and anti- government. In particular, we computed the number of comments and their gender attribution on a monthly basis based on the video’s release date (regardless of the comment’s writing date). In addition, for each month, we computed the most common words, divided by gender and political agenda. Importantly, we removed any stop words in order to obtain meaningful words rather than linguistic-related ones.

%\footnote{In natural language processing (NLP), stop words refer to commonly used words that are often filtered out during text preprocessing to improve computational efficiency and focus on more meaningful words. These stop words typically include high-frequency words like conjunctions (e.g., "and," "but," "or"), prepositions (e.g., "in," "on," "at"), and some common verbs (e.g., "is," "are," "was").} 

\section{Results}
Overall, we collected data from 2,168 videos with around 64\% (1,390) being anti-government while the remaining 36\% (778) are pro-government. On average, each video has 28.3 comments with a 108.1 standard deviation. 

Table \ref{table:1} presents the pro- and anti- government distributions of the comment, as an average of the entire two-year period. The unknown gender corresponds to users that their gender was not able to detect with at least 90\% confidence from both their user’s name and textual style. One can notice that this value is around 30\% of the data, leaving us with about 70\% of the comments with a gender identity. Moreover, the pro-government distribution clearly differed from the anti-government one in terms of gender distribution, as females are more likely to comment on anti-government channels. 

\begin{table}[!ht]
\centering
\begin{tabular}{ccc}
\hline \hline
\textbf{Political perspective} & \textbf{Gender} & \textbf{Percentage} \\ \hline \hline
 & Unknown & 27.32\% \\
Pro-government & Male & 53.14\% \\
 & Female & 19.54\% \\ \hline
& Unknown & 31.95\% \\
Anti-government  & Male & 29.71\% \\
 & Female & 38.34\% \\ \hline \hline
\end{tabular}
\caption{The average percentage of comments over the period February 2022 and February 2024, divided into gender and political perspectives.}
\label{table:1}
\end{table}

\begin{figure}[!ht]
    \centering
    \includegraphics[width=0.99\textwidth]{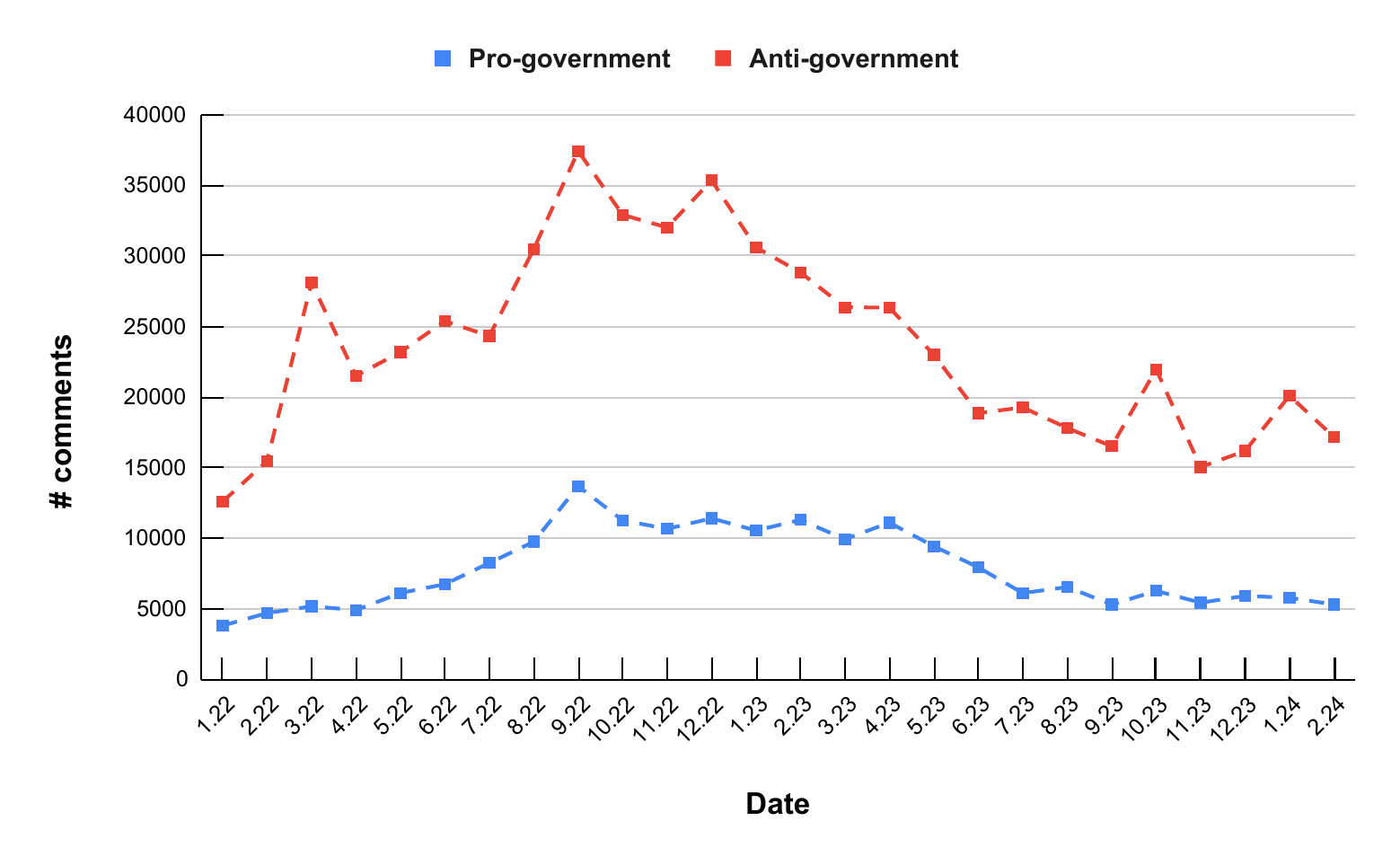}
    \caption{ The total amount of comments for each group of channels, across all their videos in a monthly aggregated manner. The dotted (red) line indicates the anti-government channels while the squared (blue) line indicates the pro-government channels. }
    \label{fig:1}
\end{figure}

Figure \ref{fig:1} shows the total amount of comments for each group of channels, across all their videos in a monthly aggregated manner. One can notice a drastic increase in March 2022 which occurs slightly after the beginning of the Russian-Ukranian war. Another peak in the anti-government activity appeared around September of the same year, which can be associated with the increasing discussion about army mobilization. Afterward, a slow but steady decrease in activity is present. Similar patterns are present for the pro-government channels, except for the initial anomaly activity around March 2022. Moreover, one can notice that after almost two years, the amount of comments, for both pro- and anti- government, returned to its pre-war volume. 

\begin{figure}[!ht]
    \centering
    \includegraphics[width=0.99\textwidth]{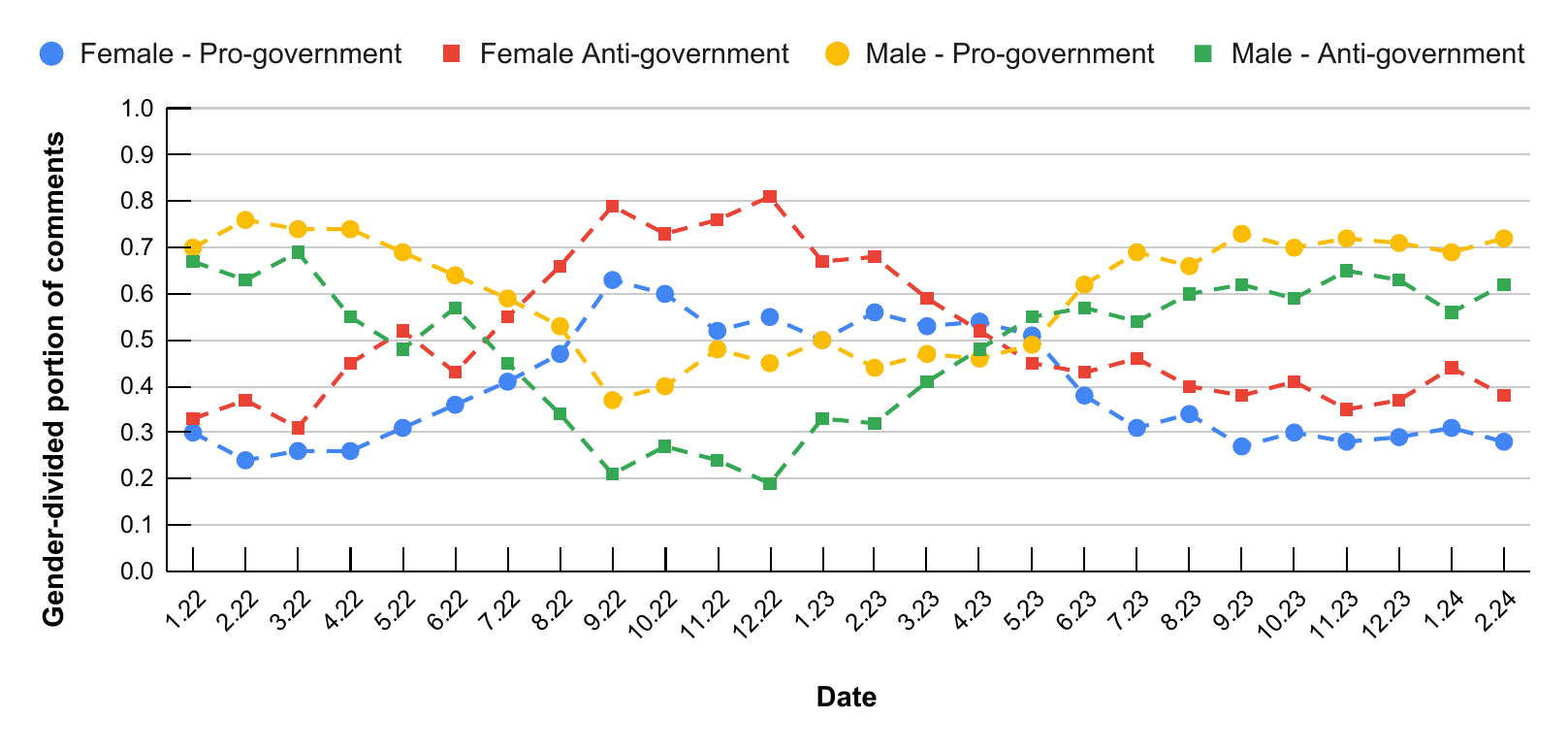}
    \caption{The portion of comments written by each gender (woman and man), divided into pro- and anti-government channels. }
    \label{fig:2}
\end{figure}

Figure \ref{fig:2} displays the portion of comments written by each gender, divided into pro- and anti- government channels. Right after the beginning of the Russian-Ukrainian war, the relative portion of women, for both pro- and anti- government, is increasing to September 2022. Following that, a steady but slow decrease in the women's portion is present until May of 2023 when a sharper decline occurs. Following this point, the gender-associated portion of the comments is kept similar to those of the pre-war ones.   

Table \ref{table:2} shows the most common word for each gender, divided into pro- and anti-government channels, over time. In order to ensure non-trend words are captured. The Table reveals notable differences in the most common words used by male and female commenters on Russian YouTube political channels. Female commenters tend to use a more diverse range of terms, often employing emotionally charged words like "shame" and "corruption," particularly in anti-government channels. They also focus more on specific domestic issues, economic concerns, and opposition figures like Navalny. In contrast, male commenters' language appears more consistent over time, frequently focusing on broader geopolitical terms and consistently mentioning Putin. While both genders use war-related terminology, males tend to use "war" more consistently, whereas females show a shift to terms like "operation" or "negotiation" in certain periods. These differences suggest varying priorities and engagement styles between genders in online political discourse. With that it's important to note that these are broad trends based on single words and don't capture the full context or individual variations within each gender group.

\begin{table}[!ht]
\centering
\begin{tabular}{c|cc|cc}
\hline \hline
\multirow{2}{*}{\textbf{\begin{tabular}[c]{@{}c@{}}Time\\ (month.year)\end{tabular}}} & \multicolumn{2}{c|}{\textbf{Anti-government}} & \multicolumn{2}{c}{\textbf{Pro-government}} \\
 & \textbf{Male} & \textbf{Female} & \textbf{Male} & \textbf{Female} \\
\hline \hline
01.22 & USA & year & Putin & Putin \\
02.22 & war & war & war & Ukraine \\
03.22 & war & corruption & Putin & war \\
04.22 & Ukraine & repression & repression & Ukraine \\
05.22 & protest & protest & operation  & protest \\
06.22 & Putin & Putin & Putin & Putin\\
07.22 & Putin & shame  & war & mobilization  \\
08.22 & Putin & shame & operation & repression \\
09.22 & corruption & corruption & NATO & mobilization  \\
10.22 & economy & security & work & mobilization \\
11.22 & Putin & Navalny & media & NATO \\
12.22 & Ukraine & Ukraine & Putin & protest \\
01.23 & work & corruption & China & year \\
02.23 & USA & Ukraine & USA & USA \\
03.23 & USA & Russia & USA & Russia \\
04.23 & China & populism & war & war \\
05.23 & war & war & operation & war  \\
06.23 & economy & rights & war & repression \\
07.23 & sanctions & propaganda & negotiations & propaganda \\
08.23 & Putin & territorial & sanctions & sanctions \\
09.23 & humanitarian  & propaganda & Putin & Putin \\
10.23 & protest & corruption & refugees & law \\
11.23 & Hague & Hague & people & peace  \\
12.23 & Putin & Navalny & Putin & escalation \\
01.24 & Navalny & Navalny & Navalny & Navalny \\
02.24 & election & propaganda & election & media \\ 
\hline \hline 
\end{tabular}
\caption{The most popular word for each month over the period February 2022 and February 2024, divided by anti- and pro- government as well as according to their gender.}
\label{table:2}
\end{table}

\section{Discussion and Conclusion}

This study explores the intricate dynamics of online political participation within the context of Russia, particularly focusing on the impact of the Russian-Ukrainian war on virtual discourse by focusing on comments written on Youtube political channels. We employed a systematic approach to data collection, content analysis, and statistical analysis, leveraging recent AI technology. This rigorous methodology facilitated the exploration of gender distribution, commenting trends, and prevalent themes within the online discourse.

The results unveiled several key findings. Initially, significant disparities in gender distribution were observed between pro- and anti-government channels, reflecting distinct patterns of participation within different segments of the online community, as indicated by Figure \ref{fig:1}. Additionally, temporal trends indicated fluctuations in commenting activity, particularly in response to pivotal events like the onset of the Russian-Ukrainian war and the great mobilization occurring during September 2022 are reflected in the data as sharp increases in activity. Second, our results show that in the context of wartime where men are called to duty, women operate as the social engine at home which is also reflected by their increased activity online compared to the manly which is higher in times of peace, as reflected by Fig. \ref{fig:2}.

These findings have several implications for our understanding of online political participation in authoritarian contexts. YouTube emerges as a significant arena for political discourse in Russia, suggesting that video-based platforms may offer unique affordances for political expression and engagement. The increased participation of women during conflict periods challenges traditional assumptions about gender dynamics in political engagement and warrants further exploration. The clear correlation between real-world events and online activity patterns demonstrates the potential of digital platforms as sensitive indicators of public sentiment. Additionally, the changing prominence of certain keywords over time reflects the dynamic nature of political discourse and the potential for online platforms to capture shifts in public concerns and priorities.

Future research could build upon these findings by conducting in-depth qualitative analysis of comment content to better understand the nuances of political expression, expanding the study to include other social media platforms for a more comprehensive view of online political discourse in Russia, investigating the potential causal relationships between online discourse patterns and offline political behaviors or attitudes, and exploring the role of platform-specific features (e.g., YouTube's recommendation algorithms) in shaping political discourse and participation.

In conclusion, this study provides valuable insights into the complex dynamics of online political participation in Russia during a period of significant geopolitical tension. By leveraging innovative methodologies and focusing on gender dynamics, our research contributes to a more nuanced understanding of digital activism and political expression in authoritarian contexts. These findings not only enhance our theoretical understanding of online political behavior but also offer practical implications for policymakers, activists, and platform developers seeking to navigate the evolving landscape of digital political engagement.

%\begin{acknowledgement}
%Elizaveta Savchenko wishes to thank Ariel University's financial support during this research.
%\end{acknowledgement}

%\paragraph{Funding Statement}
%This research did not receive any specific grant from funding agencies in the public, commercial, or not-for-profit sectors. 

%\paragraph{Competing Interests}
%None

%\endnote in some journals will behave like \footnote; and \printendnotes will not output anything. 
%\printendnotes
\clearpage
\bibliography{example}

\begin{thebibliography}{}

\bibitem[Akdeniz, 2002]{7}
Akdeniz, Y. (2002).
\newblock Anonymity, democracy, and cyberspace.
\newblock {\em Social Research: An International Quarterly}, 69:223--237.

\bibitem[Alieva, 2022]{26}
Alieva, S. (2022).
\newblock State regulation of mass communications over last decade.
\newblock pages 96--103.

\bibitem[Alyukov, 2022]{20}
Alyukov, M. (2022).
\newblock Making sense of the news in an authoritarian regime: Russian television viewers’ reception of the russia–ukraine conflict.
\newblock {\em Europe-Asia Studies}, 74(3):337--359.

\bibitem[Bail et~al., 2018]{47}
Bail, C.~A., Argyle, L.~P., Brown, T.~W., Bumpus, J.~P., Chen, H., Hunzaker, M. B.~F., Lee, J., Mann, M., Merhout, F., and Volfovsky, A. (2018).
\newblock Exposure to opposing views on social media can increase political polarization.
\newblock {\em Proceedings of the National Academy of Sciences}, 115(37):9216--9221.

\bibitem[Bakshy et~al., 2015]{48}
Bakshy, E., Messing, S., and Adamic, L. (2015).
\newblock Political science. exposure to ideologically diverse news and opinion on facebook.
\newblock {\em Science (New York, N.Y.)}, 348.

\bibitem[Bereni, 2021]{33}
Bereni, L. (2021).
\newblock The women’s cause in a field: rethinking the architecture of collective protest in the era of movement institutionalization.
\newblock {\em Social Movement Studies}, 20(2):208--223.

\bibitem[Briggs, 2008]{46}
Briggs, J.~E. (2008).
\newblock Young women and politics: an oxymoron?
\newblock {\em Journal of Youth Studies}, 11(6):579--592.

\bibitem[Cagé, 2020]{15}
Cagé, J. (2020).
\newblock Media competition, information provision and political participation: evidence from french local newspapers and elections, 1944–2014.
\newblock {\em Journal of Public Economics}, 185.

\bibitem[Carnaghan, 1996]{21b}
Carnaghan, E. (1996).
\newblock Alienation, apathy, or ambivalence? ‘don’t knows’ and democracy in russia.
\newblock {\em Slavic Review}, 55(2):325--363.

\bibitem[Chulitskaya and Matonyte, 2024]{38}
Chulitskaya, T. and Matonyte, I. (2024).
\newblock State violence and pains of punishment: Experiences of incarcerated women in belarus in the aftermath of the 2020 protests.
\newblock {\em Nationalities Papers}, pages 1--17.

\bibitem[Daphi et~al., 2023]{43}
Daphi, P., Haunss, S., Sommer, M., and Teune, S. (2023).
\newblock Taking to the streets in germany – disenchanted and confident critics in mass demonstrations.
\newblock {\em German Politics}, 32(3):440--468.

\bibitem[de~Moor et~al., 2020]{41}
de~Moor, J., Uba, K., Wahlström, M., Wennerhag, M., and De~Vydt, M., editors (2020).
\newblock {\em Protest for a future II: Composition, mobilization and motives of the participants in Fridays For Future climate protests on 20-27 September, 2019, in 19 cities around the world}.

\bibitem[Enikolopov et~al., 2020]{16}
Enikolopov, R., Makarin, A., and Petrova, M. (2020).
\newblock Social media and protest participation: Evidence from russia.
\newblock {\em Econometrica}, 88:1479--1514.

\bibitem[Gaffney, 2010]{14b}
Gaffney, D. (2010).
\newblock Iran election: quantifying online activism.
\newblock In {\em Proceedings of WebSci'10: Extending the Frontiers of Society On-Line}, Raleigh, NC, USA.

\bibitem[García et~al., 2020]{selenium}
García, B., Gallego, M., Gortazar, F., and Munoz-Organero, M. (2020).
\newblock A survey of the selenium ecosystem.
\newblock {\em Electronics}, 9(7).

\bibitem[Gel'man, 2021]{31}
Gel'man, V. (2021).
\newblock Constitution, authoritarianism, and bad governance: The case of russia.
\newblock {\em Russian Politics}, 6(1):71--90.

\bibitem[Ghonim, 2012]{10}
Ghonim, W. (2012).
\newblock {\em Revolution 2.0: The Power of the People Is Greater Than the People in Power: A Memoir}.
\newblock Houghton Mifflin Harcourt.

\bibitem[Golosov, 2023]{30}
Golosov, G. (2023).
\newblock The place of russia’s political regime (2003–2023) on a conceptual map of the world’s autocracies.
\newblock {\em Social Science Information}, 62(3):390--408.

\bibitem[Grasso and Smith, 2022]{36}
Grasso, M. and Smith, K. (2022).
\newblock Gender inequalities in political participation and political engagement among young people in europe: Are young women less politically engaged than young men?
\newblock {\em Politics}, 42(1):39--57.

\bibitem[Guriev et~al., 2021]{13}
Guriev, S., Melnikov, N., and Zhuravskaya, E. (2021).
\newblock 3g internet and confidence in government.
\newblock {\em The Quarterly Journal of Economics}, 136(4):2533--2613.

\bibitem[Haider, 2009]{5}
Haider, A. (2009).
\newblock Contribution of internet to a democratic society.
\newblock {\em ECIS 2009 Proceedings}, pages~--.

\bibitem[Halberstam and Knight, 2016]{50}
Halberstam, Y. and Knight, B. (2016).
\newblock Homophily, group size, and the diffusion of political information in social networks: Evidence from twitter.
\newblock {\em Journal of Public Economics}.

\bibitem[Hampton et~al., 2017]{51}
Hampton, K.~N., Shin, I., and Lu, W. (2017).
\newblock Social media and political discussion: when online presence silences offline conversation.
\newblock {\em Information, Communication \& Society}, 20(7):1090--1107.

\bibitem[Hensby, 2021]{21}
Hensby, A. (2021).
\newblock Political non-participation in elections, civic life and social movements.
\newblock {\em Sociology Compass}, 15.

\bibitem[Herasimenka, 2022]{18}
Herasimenka, A. (2022).
\newblock Movement leadership and messaging platforms in preemptive repressive settings: Telegram and the navalny movement in russia.
\newblock {\em Social Media + Society}, 8(3).

\bibitem[Hessami and Fonseca, 2020]{41a}
Hessami, Z. and Fonseca, M. (2020).
\newblock Female political representation and substantive effects on policies: A literature review.
\newblock {\em European Journal of Political Economy}, 63.

\bibitem[Hu et~al., 2021]{gender_model}
Hu, Y., Hu, C., Tran, T., Kasturi, T., Joseph, E., and Gillingham, M. (2021).
\newblock What’s in a name? – gender classification of names with character based machine learning models.
\newblock {\em Data Mining and Knowledge Discovery}, 4.

\bibitem[Jardine, 2015]{4}
Jardine, E. (2015).
\newblock The dark web dilemma: Tor, anonymity and online policing.
\newblock {\em Global Commission on Internet Governance}, pages~--.

\bibitem[Kadiwal, 2021]{35}
Kadiwal, L. (2021).
\newblock Feminists against fascism: The indian female muslim protest in india.
\newblock {\em Education Sciences}, 11(12):793.

\bibitem[Kaye, 2022]{26a}
Kaye, D. (2022).
\newblock Online propaganda, censorship and human rights in russia’s war against reality.
\newblock {\em AJIL Unbound}, 116:140--144.

\bibitem[Kukshinov, 2021]{22}
Kukshinov, E. (2021).
\newblock Discourse of non-participation in russian political culture: Analyzing multiple sites of hegemony production.
\newblock {\em Discourse \& Communication}, 15(2):163--183.

\bibitem[Levy and Razin, 2019]{49}
Levy, G. and Razin, R. (2019).
\newblock Echo chambers and their effects on economic and political outcomes.
\newblock {\em Annual Review of Economics}, 11:303--328.

\bibitem[Litvinenko, 2021]{19}
Litvinenko, A. (2021).
\newblock Youtube as alternative television in russia: Political videos during the presidential election campaign 2018.
\newblock {\em Social Media + Society}, 7(1).

\bibitem[Liu, 2022]{34}
Liu, S.-J.~A. (2022).
\newblock Gender gaps in political participation in asia.
\newblock {\em International Political Science Review}, 43(2):209--225.

\bibitem[Lotan et~al., 2011]{14a}
Lotan, G., Graeff, E., Ananny, M., Gaffney, D., and Pearce, I. (2011).
\newblock The arab spring. the revolutions were tweeted: Information flows during the 2011 tunisian and egyptian revolutions.
\newblock {\em International Journal of Communication}, 5.

\bibitem[Mercea, 2022]{42}
Mercea, D. (2022).
\newblock Tying transnational activism to national protest: Facebook event pages in the 2017 romanian \#rezist demonstrations.
\newblock {\em New Media \& Society}, 24(8):1771--1790.

\bibitem[Morozov, 2011]{11}
Morozov, E. (2011).
\newblock {\em The Net Delusion: The Dark Side of Internet Freedom}.
\newblock Perseus, New York.

\bibitem[Mosquera et~al., 2020]{12}
Mosquera, R., Odunowo, M., McNamara, T., Guo, X., and Petrie, R. (2020).
\newblock The economic effects of facebook.
\newblock {\em Experimental Economics}, 23:575--602.

\bibitem[Neundorf and Pop-Eleches, 2020]{28}
Neundorf, A. and Pop-Eleches, G. (2020).
\newblock Dictators and their subjects: Authoritarian attitudinal effects and legacies.
\newblock {\em Comparative Political Studies}, 53(12):1839--1860.

\bibitem[Nkansah, 2022]{39}
Nkansah, G.~B. (2022).
\newblock Youth cohort size, structural socioeconomic conditions, and youth protest behavior in democratic societies (1995–2014).
\newblock {\em Sage Open}, 12(2).

\bibitem[Otter et~al., 2021]{dl_in_nlp}
Otter, D.~W., Medina, J.~R., and Kalita, J.~K. (2021).
\newblock A survey of the usages of deep learning for natural language processing.
\newblock {\em IEEE Transactions on Neural Networks and Learning Systems}, 32(2):604--624.

\bibitem[Pirannejad, 2017]{2}
Pirannejad, A. (2017).
\newblock Can the internet promote democracy? a cross-country study based on dynamic panel data models.
\newblock {\em Information Technology for Development}, 23(2):281--295.

\bibitem[Prokop and Hrehorowicz, 2019]{21a}
Prokop, M. and Hrehorowicz, A. (2019).
\newblock Between political apathy and political passivity. the case of modern russian society.
\newblock {\em Torun International Studies}, 1:109.

\bibitem[Rosenfeld, 2017]{37}
Rosenfeld, B. (2017).
\newblock Reevaluating the middle-class protest paradigm: A case-control study of democratic protest coalitions in russia.
\newblock {\em American Political Science Review}, 111(4):637--652.

\bibitem[Saunders and Shlomo, 2021]{32}
Saunders, C. and Shlomo, N. (2021).
\newblock A new approach to assess the normalization of differential rates of protest participation.
\newblock {\em Qualitative and Quantitative}, 55:79--102.

\bibitem[Sawyer et~al., 2022]{44}
Sawyer, P.~S., Romanov, D.~M., Slav, M., and Korotayev, A.~V. (2022).
\newblock Urbanization, the youth, and protest: A cross-national analysis.
\newblock {\em Cross-Cultural Research}, 56(2-3):125--149.

\bibitem[Srinath, 2017]{python}
Srinath, K.~R. (2017).
\newblock Python – the fastest growing programming language.
\newblock {\em International Research Journal of Engineering and Technology}, 4(12).

\bibitem[Stieglitz and Dang-Xuan, 2013]{9}
Stieglitz, S. and Dang-Xuan, L. (2013).
\newblock Social media and political communication: a social media analytics framework.
\newblock {\em Social Network Analysis and Mining}, 3:1277--1291.

\bibitem[Tannenberg, 2022]{27}
Tannenberg, M. (2022).
\newblock The autocratic bias: self-censorship of regime support.
\newblock {\em Democratization}, 29(4):591--610.

\bibitem[Temir, 2021]{17}
Temir, E. (2021).
\newblock Power, opposition and social media in russia.
\newblock {\em Gümüşhane Üniversitesi İletişim Fakültesi Elektronik Dergisi}, 9(1):470--501.

\bibitem[Toepfl, 2018]{14}
Toepfl, F. (2018).
\newblock From connective to collective action: internet elections as a digital tool to centralize and formalize protest in russia.
\newblock {\em Information, Communication \& Society}, 21(4):531--547.

\bibitem[Tufekci and Wilson, 2012]{8}
Tufekci, Z. and Wilson, C. (2012).
\newblock Social media and the decision to participate in political protest: Observations from tahrir square.
\newblock {\em Journal of Communication}, 62(2):363--379.

\bibitem[Weinstein, 2014]{45}
Weinstein, E. (2014).
\newblock The personal is political on social media: Online civic expression patterns and pathways among civically engaged youth.
\newblock {\em International Journal of Communication}, 8:24.

\bibitem[Zhelnina, 2020]{23}
Zhelnina, A. (2020).
\newblock The apathy syndrome: How we are trained not to care about politics.
\newblock {\em Social Problems}, 67(2):358--378.

\bibitem[Zhuravlev et~al., 2020]{53}
Zhuravlev, O., Savelyeva, N., and Erpyleva, S. (2020).
\newblock The cultural pragmatics of an event: the politicization of local activism in russia.
\newblock {\em International Journal of Politics, Culture, and Society}, 33:163–180.

\bibitem[Zhuravskaya et~al., 2020]{1}
Zhuravskaya, E., Petrova, M., and Enikolopov, R. (2020).
\newblock Political effects of the internet and social media.
\newblock {\em Annual Review of Economics}, 12.

\end{thebibliography}
\bibliographystyle{apalike}

\appendix
\setcounter{table}{0}
\renewcommand{\thetable}{A\arabic{table}}
\setcounter{figure}{0}
\renewcommand{\thefigure}{A\arabic{figure}}
%\section{Data sources}
%Table \ref{table:appendix} outlines the YouTube channels considered for this study. 

\begin{table}[!ht]
\begin{tabular}{ll}
\hline \hline
\textbf{Political perspective} & \textbf{Source} \\ \hline \hline
Pro-government & \begin{tabular}[c]{@{}l@{}}https://youtube.com/@aifru?si=ZSAgaa7vM-08FDPC\\ https://youtube.com/@Ugratv1?si=U3B9wtjD8TP-VyJY\\ https://youtube.com/@kpru?si=2ubxTblqYvsOxCp3\\ https://youtube.com/@ua2378?si=KOZUgPp1I7Q\_HBfz\\ https://youtube.com/@rogandar?si=7kMhogj2lhVX06Fu\\ https://youtube.com/@SlavaYa\_101?si=KFFEuQTTpYVwdB4h\\ https://youtube.com/@POYMI?si=gxYCYq9nITH7nxj9\\ https://youtube.com/@Vestnik\_Buri?si=FrTi45gP46V-MsJG\\ https://youtube.com/@moscowfm?si=8UsmveLVt-\_Gy\_BL\\ https://youtube.com/@user-bg8db3xk1e?si=O49UOkWarNqesuCi\\ https://youtube.com/@walkandtalk\_?si=DlYKramzzNrPb4AJ\\ https://youtube.com/@sobchak?si=rr04rMlBfH28dfPI\\ https://youtube.com/@Diplomatrutube?si=kHSk2IcTTWe-KoHm\\ https://youtube.com/@svoysredisvoih?si=\_iKQdvJoWvee8h4N\\ https://youtube.com/@sovetskij?si=UL4xqcicdYZx5A3b\\ https://rutube.ru/u/solovievlive/\\ https://rutube.ru/channel/23174740/\\ https://rutube.ru/u/sputnik/\\ https://rutube.ru/u/ntv/\\ https://rutube.ru/channel/4478643/\end{tabular} \\ \hline
Anti-government & \begin{tabular}[c]{@{}l@{}}https://youtube.com/@holodmedia?si=0Saj-XqqWm6Dc16I\\ https://youtube.com/@redactsiya?si=jU0evWCKs-8ePbpZ\\ https://youtube.com/@NavalnyRu?si=\_mPZSk0rNev8tC0n\\ https://youtube.com/@vdud?si=6ylR1VvNEZCr0QZm\\ https://youtube.com/@tvrain?si=1EKB0fEKXArJJAC4\\ https://youtube.com/@zhivoygvozd?si=WnlOgxS40mzIN3PP\\ https://youtube.com/@khodorkovskylive?si=jLgcuOqC3r8ivpMl\\ https://youtube.com/@Radio-Svoboda?si=3e00ZevA0z5B\_dQK\\ https://youtube.com/@bbcnewsrussian?si=FJ-HVJMgsZ8IS3Lf\\ https://youtube.com/@dwrussian?si=s82fuhj5gDGxEeUg\\ https://youtube.com/@mediazzzona?si=FY3eNYk2WOuspeBa\\ https://youtube.com/@videogolos?si=6WL-wbVb8u2YsFvf\\ https://youtube.com/@publicverdict?si=EvPvZFG7ckkzsysw\\ https://youtube.com/@MemoRu?si=UwH1cas77QF-vIZv\\ https://youtube.com/@Max\_Katz?si=60gbpAYp8oTIP7jN\\ https://youtube.com/@no\_torture?si=ayCtruphyzct5cyq\\ https://youtube.com/@SVTVofficial?si=r4dytHVloHQNFquF\\ https://youtube.com/@Ekaterina\_Schulmann?si=Kpxiwe4QHHwNykpz\\ https://youtube.com/@MeduzaPro?si=Ml20TiWuPctlmfn3\\ https://youtube.com/@yashin\_russia?si=bhLE00IuhVQjWvLA\\ https://youtube.com/@proekt\_media?si=2gfJMgxuThlJQlbL\\ https://youtube.com/@istories\_media?si=0ZulUv8\_tNnuRTS7\end{tabular} \\ \hline \hline
\end{tabular}
\caption{The channels used for this study, are divided into pro- and anti- government.}
\label{table:appendix}
\end{table}

\end{document}